\documentclass[
  aps,prd,reprint,twocolumn,nofootinbib,
  amsmath,amssymb,floatfix,preprintnumbers,longbibliography
]{revtex4-2}

\usepackage{amsmath,amssymb,bm,mathtools}
\usepackage{graphicx}
\usepackage{xcolor}
\usepackage{hyperref}
\hypersetup{
  colorlinks=true,
  linkcolor=blue,
  citecolor=blue,
  urlcolor=blue
}

\newcommand{\ii}{\mathrm i}
\newcommand{\ee}{\mathrm e}
\newcommand{\tr}{\operatorname{tr}}

\renewcommand{\Im}{\operatorname{Im}}

\begin{document}

\preprint{RESCEU-27/26}

\title{Fractal phase structure of QCD under imaginary rotation}

\author{Rin Takada}
\email{takada-rin@resceu.s.u-tokyo.ac.jp}
\affiliation{Research Center for the Early Universe (RESCEU), Graduate School of Science, The University of Tokyo, 7-3-1 Hongo, Bunkyo, Tokyo 113-0033, Japan}
\date{\today}

\begin{abstract}
We show that in hot QCD rotating at an imaginary angular velocity $\tilde\Omega_I$, the angular velocity selects not only the temperature of the bulk but whether charge conjugation $C$ is broken there.
For QCD at temperature $T$ and imaginary quark chemical potential $\theta$, with $\tilde\Omega_I/2\pi=p/q$ in lowest terms, the bulk far from the rotation axis is in the same thermal equilibrium state as nonrotating QCD at temperature $T/q$ and imaginary quark chemical potential $\theta'=q\theta+\pi(p+q+1)$; this follows from the Euclidean boundary conditions and locality alone, and holds nonperturbatively whenever the equivalent nonrotating system has a finite correlation length.
Consequently, even at zero quark chemical potential, the bulk is placed at the Roberge-Weiss (RW) point whenever $p$ and $q$ are both odd, and $C$ is spontaneously broken there for $T>qT_{\rm RW}$, with $T_{\rm RW}$ the RW endpoint temperature.
The bulks at $\tilde\Omega_I=2\pi/3$ and $4\pi/3$, for example, are both at temperature $T/3$, yet only the former can break $C$.
The set of imaginary angular velocities at which $C$ is broken has a fractal structure: raising $T$ adds rationals of ever larger denominator.
At irrational $\tilde\Omega_I/2\pi$ the bulk corresponds to zero-temperature QCD at any $T$ and remains confined.
\end{abstract}

\maketitle

\section{Introduction}
The bulk of a thermal system rotating at an imaginary angular velocity
$\tilde\Omega_I$ depends on that angular velocity in an unusual way.
For free fields at temperature $T$, when $\tilde\Omega_I/(2\pi)$ is a fraction $p/q$ in lowest terms, the free energy density far from the axis is that of the nonrotating theory at temperature $T/q$, with the fermion boundary condition fixed by the parity of $p+q$, and when $\tilde\Omega_I/(2\pi)$ is irrational, it is the zero-temperature value~\cite{Chernodub:2022wsx,Ambrus:2023bid,Patuleanu:2025yfj}; thermal effective field theory extends the rational case to interacting theories~\cite{Benjamin:2024rvn}.
The bulk thermodynamics thus changes discontinuously as
$\tilde\Omega_I$ is varied continuously.
Which symmetry the parity of $p+q$ concerns, and what it does to the phases of QCD at temperature $T/q$, are the questions we address.

Our starting point is what is known about QCD at imaginary angular velocity.
It has a real Euclidean action, so it can be simulated directly on the lattice~\cite{Yamamoto:2013zwa,Braguta:2020biu,Braguta:2021jgn,Chernodub:2022veq,Braguta:2023yjn}; its phase structure has also been studied at one loop~\cite{Chen:2022smf,Chen:2024tkr,Takada:2026rw}.
Using an identity on the color-singlet space, we showed that QCD rotating at imaginary angular velocity $2\pi$ maps exactly onto nonrotating QCD at imaginary quark chemical potential $\pi$, which is the Roberge-Weiss (RW) point~\cite{Takada:2026rw}.
At the RW point, charge conjugation $C$ is spontaneously broken above the RW endpoint temperature $T_{\rm RW}$~\cite{Roberge:1986mk,DElia:2009pdy,Bonati:2016pwz}.
We also showed that in the one-loop analysis in the high-temperature limit with three massless flavors, $C$ breaking on the axis occupies a window $1.40\pi\lesssim\tilde\Omega_I\lesssim 2.60\pi$ around
$\tilde\Omega_I=2\pi$~\cite{Takada:2026rw}.
Below we settle what this implies for the bulk.

QCD has the deconfinement and chiral crossovers and, at imaginary quark chemical potential, the RW transition, each with its own characteristic temperature; if the bulk behaves as a system at temperature $T/q$, these temperatures should appear multiplied by $q$.
In this paper we show that in imaginarily rotating QCD the imaginary angular velocity sets not only the temperature of the bulk but also its imaginary quark chemical potential, and hence whether $C$ can be broken there.
The bulks at $\tilde\Omega_I=2\pi/3$ and $4\pi/3$, for example, are both at temperature $T/3$, but their imaginary quark chemical potentials are $\pi$, the RW point, and $0$, so that with rotation alone $C$ is spontaneously broken above $3T_{\rm RW}$ in the former bulk but not in the latter.

In QCD at imaginary angular velocity $\tilde\Omega_I=2\pi p/q$ and imaginary quark chemical potential $\theta$, the bulk is, nonperturbatively, in the same thermal equilibrium state as nonrotating QCD at temperature $T/q$ and imaginary chemical potential $\theta'=q\theta+\pi(p+q+1)$, which we call the equivalent system.
Its state is probed not by the ordinary Polyakov loop used on the axis but by the imaginary part of a $q$-fold Polyakov loop, which closes only after passing through a point and its $q-1$ images under the rotation, and by the imaginary quark number density.
The resulting phase structure can be tested on the lattice free of the sign problem.

\begin{figure*}[t]
\centering
\includegraphics[width=.9\textwidth]{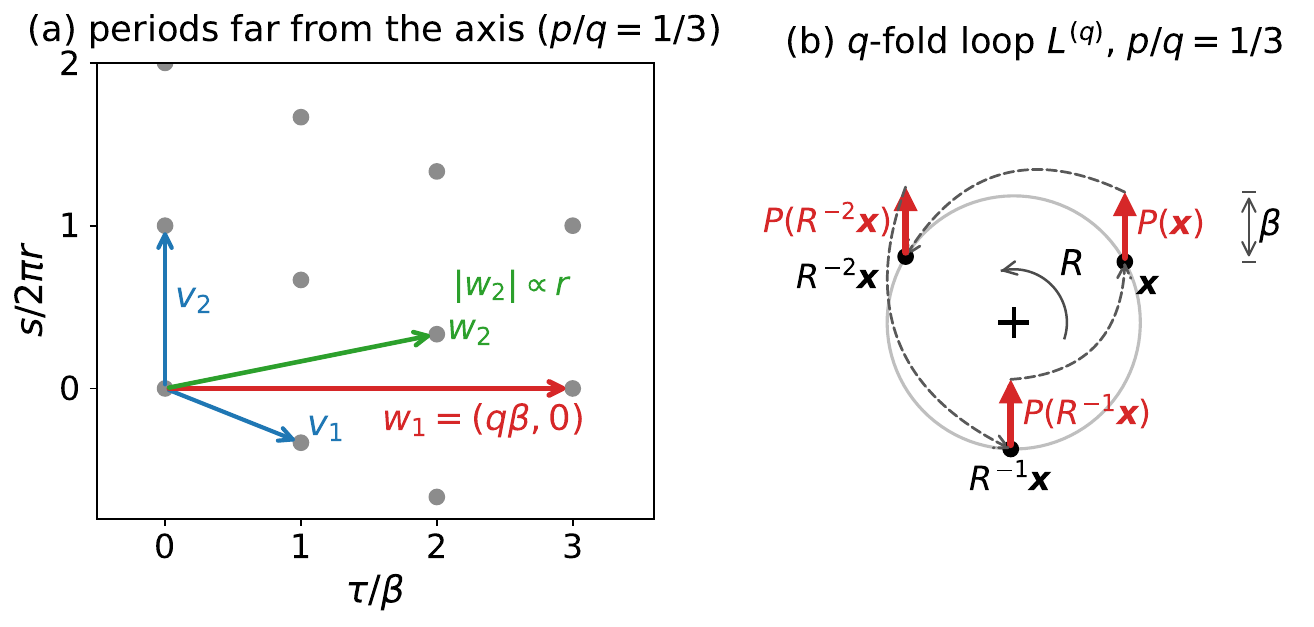}
\caption{The case $\tilde\Omega_I/2\pi=1/3$.
(a) The boundary condition far from the axis in the local coordinates
$(\tau,s=r\varphi)$; the gray dots are the same point as the origin.
The displacements generated by $v_1$ and $v_2$ include the pure imaginary-time translation $w_1=(q\beta,0)$, while the independent displacement $w_2$ grows with $r$ and drops out of the local physics far from the axis; the $s$ component of $w_2$ in the figure, $2\pi r/q$, is an arc length along the circle of radius $r$, and the distance in three-dimensional space between the two points connected by $w_2$ is the chord $2r\sin(\pi/q)$.
(b) The $q$-fold Polyakov loop $L^{(q)}$ of Eq.~\eqref{eq:Lq}.
The red arrows are the Wilson lines $P(R^{-k}\bm x)$ ($k=0,1,2$) of length $\beta$, each running by the boundary condition from $R^{-k}\bm x$ to $R^{-(k-1)}\bm x$, with $R$ the rotation matrix by $2\pi p/q$ about the axis; the dashed lines show the identification that joins the end point of
$P(R^{-k}\bm x)$ to the starting point of $P(R^{-(k-1)}\bm x)$.}
\label{fig:geom}
\end{figure*}

\section{Bulk equivalence and order parameters}
In hot $SU(3)$ QCD rotating at imaginary angular velocity
$\tilde\Omega_I$, one imposes the boundary condition
$(\tau,\varphi,r,z)\sim(\tau+\beta,\varphi-\tilde\Omega_I,r,z)$ in Euclidean space~\cite{Chen:2022smf,Chernodub:2022wsx}.
Here $T$ is the temperature on the axis, $\beta=1/T$,
$\tilde\Omega_I\coloneqq\Omega_I/T$ is the dimensionless imaginary angular velocity, and the imaginary quark chemical potential $\theta$ is defined by $\ii\theta\coloneqq\mu/T$.
Each advance of $\beta$ in imaginary time thus rotates space by $\tilde\Omega_I$.
Consider first $\tilde\Omega_I=2\pi/3$ and $4\pi/3$.
In both cases three applications of the boundary condition return to the original spatial point, carrying space once around in the former case and twice in the latter, so the pure imaginary-time translation that leaves the point fixed has length $3\beta$; one or two applications lead to spatially distinct points separated by the chord $\sqrt{3}\,r$.
A boundary condition ties together the field values at two separated points, and its effect on the local physics decays with their separation on the scale of the correlation length; we refer to this property as locality.
Far from the axis the only surviving boundary condition is therefore the imaginary-time period $3\beta$, and the bulk is at temperature $T/3$ in both cases.
The difference appears in the quark boundary condition.
In both cases the quark field acquires $(-1)^3=-1$ from antiperiodicity, but a spinor changes sign under every $2\pi$ rotation of space, so at $\theta=0$, where the imaginary chemical potential adds no phase, the total factor over the circuit of length $3\beta$ is $+1$ at $\tilde\Omega_I=2\pi/3$ and $-1$ at $4\pi/3$.
In nonrotating QCD at temperature $T/3$ and imaginary chemical potential $\theta'$, the same circuit multiplies the quark field by $-\ee^{-\ii\theta'}$.
Equating the two factors places the bulk at the RW point $\theta'=\pi$ for $\tilde\Omega_I=2\pi/3$ and at zero chemical potential for $4\pi/3$.

In general, for $\tilde\Omega_I/2\pi=p/q$ with $\gcd(p,q)=1$, applying the boundary condition $q$ times carries space $p$ times around and back to the original point, so the translation by $q\beta$ in imaginary time, which we denote by $w_1$, leaves the spatial point fixed.
To identify the other periods, write the boundary condition and the $2\pi$ periodicity of $\varphi$ in the local coordinates $(\tau,s=r\varphi)$ on the cylinder of radius $r$ as the periods $v_1=(\beta,-\tilde\Omega_Ir)$ and $v_2=(0,2\pi r)$ [Fig.~\ref{fig:geom}(a)].
With integers $a,b$ satisfying $qb-pa=1$ (B\'{e}zout's identity), $w_1=qv_1+pv_2=(q\beta,0)$ and $w_2=av_1+bv_2=(a\beta,2\pi r/q)$ generate the same periods as $v_1$ and $v_2$, since the change of basis has unit determinant (Appendix~\ref{app:periods}).
Every period $kw_1+lw_2$ with $l\not\equiv0\pmod q$ therefore has an $s$ component of magnitude at least $2\pi r/q$.
A point and the point connected to it by such a period lie on the circle of radius $r$ at a central angle that is a nonzero multiple of $2\pi/q$, so their distance in three-dimensional space is at least the chord $2r\sin(\pi/q)$ (Appendix~\ref{app:periods}), which grows without bound as $r\to\infty$; the periods with $l\equiv0\pmod q$ are integer combinations of $w_1$ and $v_2$ and return to the same spatial point.
If the correlation length of the equivalent system is finite, local observables depend neither on boundary conditions relating points farther apart than the correlation length nor on the axis, so as $r\to\infty$ the only boundary condition that affects the local thermodynamics is the imaginary-time periodicity with period $q\beta$, and the bulk is in thermal equilibrium at temperature $T/q$.
Over one circuit of $w_1$ the quark field acquires $(-1)^q$ from antiperiodicity, $\ee^{-\ii q\theta}$ from the imaginary chemical potential, and the spinor sign $(-1)^p$ from the $p$ turns of the rotation, so that
\begin{align}
\psi(\tau+q\beta)=(-1)^{p+q}\ee^{-\ii q\theta}\psi(\tau).
\label{eq:qbc}
\end{align}
In the nonrotating system at temperature $T/q$ the same circuit multiplies the quark field by $-\ee^{-\ii\theta'}$, whence $\ee^{-\ii\theta'}=(-1)^{p+q+1}\ee^{-\ii q\theta}$, that is,
\begin{align}
\theta'\equiv q\theta+\pi(p+q+1)\pmod{2\pi}.
\label{eq:chemical}
\end{align}

Let $f(T,\theta,\tilde\Omega_I)$ denote the free energy density of QCD matter far from the axis, with the zero-temperature vacuum energy subtracted; $f(T,\theta,0)$ is that of nonrotating QCD matter.
The Euclidean path integral is specified by the Lagrangian and the boundary conditions, and the rotation, the temperature and the imaginary chemical potential all enter through the latter, while the Lagrangian is that of ordinary QCD.
Far from the axis, the boundary conditions relevant to the local physics reduce to those of nonrotating QCD at temperature $T/q$ and, by Eq.~\eqref{eq:qbc}, imaginary chemical potential $\theta'$, so the path integral far from the axis coincides locally with that of the equivalent system.
Hence, as long as the correlation length of this system is finite,
\begin{align}
f\biggl(T,\theta,\dfrac{2\pi p}{q}\biggr)
=f\biggl(\dfrac Tq,\theta',0\biggr)
\label{eq:theorem}
\end{align}
holds nonperturbatively, with $\theta'$ given by Eq.~\eqref{eq:chemical}, and its derivatives with respect to $T$ and $\theta$ give the same correspondence for the energy density and the quark number density.
The finiteness of the correlation length presupposes that the equivalent system is in a pure phase.
We consider QCD at physical quark masses, where the pion is massive even at zero temperature, so the correlation length of the equivalent system in a pure phase is finite except at the RW endpoint.
Two situations require care: on the first-order line where the two $C$-broken phases coexist, correlations in a mixed state do not decay and locality fails, so one must select one pure phase, where they do decay, and apply Eq.~\eqref{eq:theorem} to it; at the critical point $T=qT_{\rm RW}$, where the correlation length diverges, one must fix $\tilde\Omega_I$, take $r\to\infty$ first at $T\neq qT_{\rm RW}$, and only then approach the critical point.
For $p=q=1$, Eq.~\eqref{eq:theorem} follows from the operator identity
$\hat\rho(\tilde\Omega_I+2\pi,\theta)=\hat\rho(\tilde\Omega_I,\theta+\pi)$, valid everywhere in space~\cite{Takada:2026rw}; for $q>1$, Eq.~\eqref{eq:theorem} holds only far from the axis, where the period $w_2$ has dropped out.
When $\tilde\Omega_I/2\pi=\alpha$ is irrational, no pure imaginary-time translation exists: the distance between a point and its image under $m\neq0$ applications of the boundary condition is
\begin{align}
d_m(r)^2=(m\beta)^2+4r^2\sin^2(\pi m\alpha).
\label{eq:dm}
\end{align}
For any distance $D$, $d_m(r)\geqslant|m|\beta$ gives $d_m(r)>D$ for all $|m|>D/\beta$, and for the finitely many remaining $0<|m|\leqslant D/\beta$, $\sin(\pi m\alpha)\neq0$ because $\alpha\notin\mathbb Q$, so $d_m(r)>D$ holds for these as well once $r$ is large enough.
Thus $\min_{m\neq0}d_m(r)$ diverges as $r\to\infty$: no period of finite length survives far from the axis, and by locality the bulk corresponds to zero temperature.
Unlike the free-field results~\cite{Chernodub:2022wsx,Ambrus:2023bid,Patuleanu:2025yfj}, this argument does not use the explicit form of the partition function, so it holds for interacting theories.

We now define the order parameters in the bulk.
The Polyakov loop of the equivalent system is the trace of the holonomy along the thermal circle $w_1$.
Since $w_1$ uses the boundary condition $q$ times, joining $q$ Wilson lines, each advancing imaginary time by $\beta$, yields the $q$-fold Polyakov loop
\begin{align}
L^{(q)}(\bm x)
=\dfrac13\tr\bigl[P(\bm x)P(R^{-1}\bm x)\cdots P(R^{-(q-1)}\bm x)\bigr],
\label{eq:Lq}
\end{align}
which generalizes the fourfold Polyakov loop introduced in Ref.~\cite{Chernodub:2022veq} for $\tilde\Omega_I=\pi/2$.
Here $R$ is the rotation matrix by $2\pi p/q$ about the $z$ axis, $P(\bm y)$ is the Wilson line running in the imaginary-time direction from $\bm y$ to $R\bm y$ across the boundary $\tau=\beta$ [Fig.~\ref{fig:geom}(b)], and the product $H_q(\bm x)=P(\bm x)P(R^{-1}\bm x)\cdots P(R^{-(q-1)}\bm x)$ is the holonomy of the equivalent system.
A single $P(\bm y)$ is not closed and transforms under a gauge transformation as $P(\bm y)\mapsto g(R\bm y)P(\bm y)g(\bm y)^{-1}$, but in $H_q$ the gauge transformation matrices cancel where one Wilson line joins the next, and $R^q=\bm1$ together with the cyclicity of the trace makes $L^{(q)}$ gauge invariant.
Joined end to end by the boundary condition, the $q$ Wilson lines form a single closed path winding once around the thermal circle $w_1$.
Since the path integral far from the axis coincides locally with that of the equivalent system, the distribution of $L^{(q)}$ in the bulk, and its correlation function between points whose separation is small compared with their distance from the axis, equal those of the Polyakov loop of the equivalent system.
In contrast, a single Wilson line $P$ running straight in the imaginary-time direction does not close away from the axis, so the ordinary Polyakov loop $L=\tfrac13\tr P$ is not a gauge-invariant observable there and enters only as a variable of a uniform-background calculation, in which the bulk free energy depends on $P$ only through $H_q=P^q$.
For $q>1$, therefore, the bulk is probed by $L^{(q)}$ rather than by $L$, while $L^{(1)}=L$.
Since $C$ acts as $L^{(q)}\mapsto L^{(q)\ast}$, $\Im L^{(q)}$ is an order parameter for $C$.
The other is the imaginary quark number density
$n_I\coloneqq-\ii n=T^{-1}\partial f/\partial\theta$, with $n$ the quark number density, of dimension $T^3$; differentiating Eq.~\eqref{eq:theorem} shows that the bulk $n_I$ equals that of the equivalent system itself.
We now turn the consequences of Eq.~\eqref{eq:theorem} into predictions testable on the lattice.

\begin{figure*}[t]
\centering
\includegraphics[width=\textwidth]{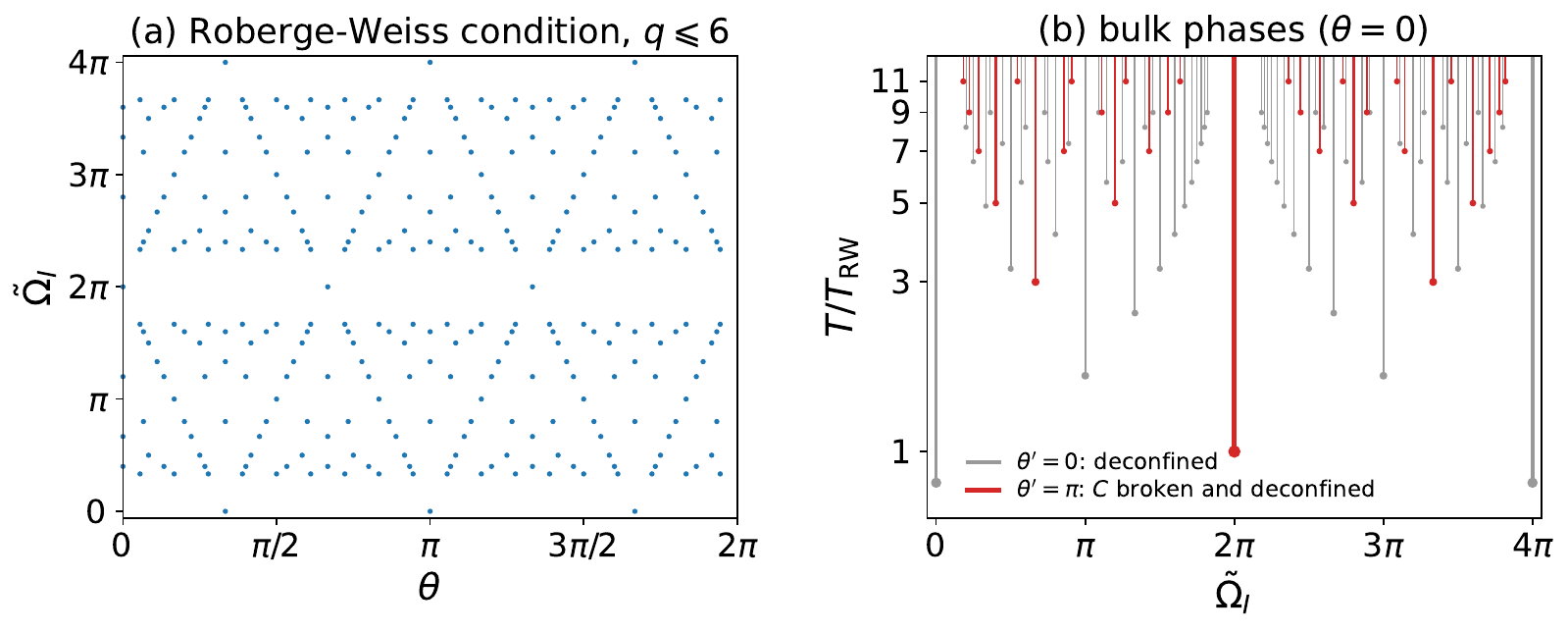}
\caption{(a) The points $(\theta,\tilde\Omega_I)$ at which the bulk lies on an RW line, for $q\leqslant6$; those with denominator $q$ are spaced by $2\pi/(3q)$ in $\theta$.
Each point is a first-order transition when $T>qT_{\rm RW}$.
(b) The state of the bulk at $\theta=0$ ($q\leqslant11$) as a function of
$\tilde\Omega_I$ and the temperature $T$ on the axis.
The red vertical lines mark the rationals $\tilde\Omega_I/2\pi=p/q$ with $p$ and $q$ both odd, for which the equivalent system is at the RW point $\theta'=\pi$; they cover the range $T>qT_{\rm RW}$ [Eq.~\eqref{eq:Cset}] in which $C$ is broken and the bulk is deconfined, and the circle at $T=qT_{\rm RW}$ marks the RW endpoint, where the chiral susceptibility is singular as well; $T_{\rm RW}=208\,\mathrm{MeV}$~\cite{Bonati:2016pwz}.
The gray vertical lines mark the rationals with $p+q$ odd, for which the equivalent system is at $\theta'=0$; the bulk is deconfined and chirally restored at high temperature, but the pseudocritical temperatures depend on the definition [Eq.~\eqref{eq:Tc}]; the circle shows $qT_c(0)$ with $T_c(0)=170\,\mathrm{MeV}$ from the inflection point of the renormalized Polyakov loop~\cite{Aoki:2009sc}; $qT_\chi(0)$ with $T_\chi(0)=156.5\,\mathrm{MeV}$~\cite{HotQCD:2018pds} lies slightly below and is not shown.}
\label{fig:comb}
\end{figure*}

\section{Phase structure of the bulk}
The free energy density $f$ of QCD matter has two symmetries in the imaginary quark chemical potential $\theta'$ of the equivalent system.
One is charge conjugation $C\colon\theta'\mapsto-\theta'$; the other is the RW periodicity $\theta'\mapsto\theta'+2\pi/3$, which holds because a shift of $\theta'$ by $2\pi/3$ can be compensated by a center transformation~\cite{Roberge:1986mk}.
Together they make $f$ invariant under the reflections
$\theta'\mapsto2\pi k/3-\theta'$ with integer $k$ about the points $\theta'\equiv\pi\pmod{2\pi/3}$, which we call the $\mathbb Z_2$ symmetry.
Above the RW endpoint temperature $T_{\rm RW}$, the two phases related by $\mathbb Z_2$ coexist on the RW lines, the first-order lines
$\theta'\equiv\pi\pmod{2\pi/3}$, on which $\mathbb Z_2$ is spontaneously broken~\cite{Roberge:1986mk,DElia:2009pdy,Bonati:2016pwz}.
On the RW line $\theta'=\pi$, $\mathbb Z_2$ is $C\colon\theta'\mapsto2\pi-\theta'\equiv-\theta'$ itself, so $C$ is spontaneously broken on this line; on the other two lines $\theta'=\pm\pi/3$, $\mathbb Z_2$ is $C$ combined with the RW periodicity, and $C$ alone is not a symmetry.
Figure~\ref{fig:comb}(a) shows the RW lines
$\theta'\equiv\pi\pmod{2\pi/3}$ in the $(\theta,\tilde\Omega_I)$ plane through Eq.~\eqref{eq:chemical}.

At $\theta=0$, Eq.~\eqref{eq:chemical} gives $\theta'=\pi$ for $p+q$ even and $\theta'=0$ for $p+q$ odd.
Since $\theta'=0$ is not on an RW line, the bulk lies on an RW line if and only if $p+q$ is even, that is, $p$ and $q$ are both odd.
For $\tilde\Omega_I=2\pi p/q$ with $p$ and $q$ both odd, once the temperature $T/q$ of the equivalent system exceeds the RW endpoint temperature,
\begin{align}
T>qT_{\rm RW},
\label{eq:Cset}
\end{align}
$C$ is spontaneously broken in the bulk, whereas at irrational imaginary angular velocities the bulk is $C$ symmetric.
At fixed temperature, only the finitely many rationals with
$q<T/T_{\rm RW}$ in one period of $\tilde\Omega_I$, $4\pi$, break $C$; as the temperature rises, rationals of ever larger denominator join in.
This structure, in which the temperature of $C$ breaking is fixed by the denominator alone, is what we call ``fractal'' in this paper.

Deconfinement in the bulk likewise follows that of the equivalent system.
At $\theta=0$ with $p+q$ odd, $\theta'=0$ and the equivalent system is ordinary QCD at zero chemical potential.
At the physical point the deconfinement and chiral crossovers~\cite{Aoki:2006we,HotQCD:2018pds}, the latter restoring the approximate chiral symmetry of the light quarks, have pseudocritical temperatures $T_c(0)$ and $T_\chi(0)$ that depend on the definition; but by Eq.~\eqref{eq:theorem} the bulk at $\theta=0$ undergoes the crossovers defined in the same way when the temperature $T/q$ of its equivalent system reaches $T_c(0)$ and $T_\chi(0)$, that is, at
\begin{align}
T=qT_c(0),\qquad T=qT_\chi(0),
\qquad(p+q\colon\text{odd}).
\label{eq:Tc}
\end{align}
When $p$ and $q$ are both odd, $\theta'=\pi$ and the equivalent system has the true criticality of the RW endpoint.
At finite quark mass the chiral condensate, being $C$-even, behaves as an energy-like operator of the RW transition, so the criticality shows up in the chiral condensate and susceptibility as well~\cite{Cuteri:2022vgo}; indeed, on the RW line the peak of the chiral susceptibility coincides with $T_{\rm RW}$ within lattice errors~\cite{Bonati:2016pwz,Cuteri:2022vgo}.
For $p$ and $q$ both odd, therefore, $T=qT_{\rm RW}$ is the exact threshold of $C$ breaking in Eq.~\eqref{eq:Cset} and a singular point of the chiral observables.

When $\tilde\Omega_I/2\pi$ is irrational, the bulk corresponds, as shown above, to nonrotating QCD at zero temperature, which at physical quark masses has a mass gap and hence a finite correlation length, so locality applies as it stands.
However high the temperature on the axis, the bulk at an irrational imaginary angular velocity therefore inherits the $C$ symmetry, the confinement and the spontaneous breaking of approximate chiral symmetry of zero-temperature QCD.
At rationals, by contrast, the bulk is deconfined and chirally restored above $qT_c(0)$ and $qT_\chi(0)$ for $p+q$ odd and above
$qT_{\rm RW}$ for $p$ and $q$ both odd, so both share the fractal structure of $C$ breaking: rationals of ever larger denominator join in as the temperature rises, while irrationals never do.
Figure~\ref{fig:comb}(b) shows this structure.

We estimate the order parameters $\Im L^{(q)}$ and $n_I$ in the
$C$-broken bulk.
At $\theta=0$ with $p$ and $q$ both odd, the equivalent system is at the RW point and $C$ is broken for $T>qT_{\rm RW}$.
When its temperature $T/q$ is high enough, the massless one-loop result at the RW point~\cite{Roberge:1986mk,Takada:2026rw}, with the temperature replaced by $T/q$, gives
\begin{align}
\Im L^{(q)}=\pm\dfrac{\sqrt3}{2},\quad
n_I=\mp\dfrac{8\pi N_f}{27}\biggl(\dfrac Tq\biggr)^3,\quad
(p,q\colon\text{odd}),
\label{eq:nI}
\end{align}
where $N_f$ is the number of flavors.
The former follows because, on the high-temperature side of the RW point, the minimum of the one-loop effective potential places the holonomy $H_q$ at the center elements $\ee^{\pm2\pi\ii/3}\bm1$, so that
$L^{(q)}=\frac13\tr H_q=\ee^{\pm2\pi\ii/3}$~\cite{Takada:2026rw}.
The two signs correspond to the two $C$-conjugate phases, selected by taking $\theta\to0^{\pm}$ after the thermodynamic limit; $C\colon\theta\mapsto-\theta$ exchanges them.
Since the bulk $n_I$ is that of the equivalent system at temperature $T/q$, the one-loop contribution from quarks heavier than that temperature, $m\gg T/q$, is suppressed by the Boltzmann factor $\ee^{-m/(T/q)}=\ee^{-qm/T}$ and decreases exponentially with the denominator $q$.

\section{Lattice tests}
Equation~\eqref{eq:Cset} is a statement about the bulk at $r\to\infty$; we do not compute finite-size corrections to it, but locality tells where on a finite lattice the bulk prediction can be probed (Appendix~\ref{app:window}), and the test does not require scanning many rationals.
The most direct test is a comparison of two imaginary angular velocities at $q=3$.
Prepare two rotating systems at temperature $T=1/(N_\tau a)$,
\begin{align}
A\colon(\tilde\Omega_I,\theta)=(2\pi/3,0),\qquad
B\colon(\tilde\Omega_I,\theta)=(4\pi/3,0),
\label{eq:pair}
\end{align}
sharing the number $N_\tau$ of temporal sites, the lattice action, the lattice spacing $a$ and the bare quark masses.
Both bulks are at temperature $T/3$, but the equivalent system of $A$ is at the RW point $\theta'=\pi$ and that of $B$ at $\theta'=0$, so Eq.~\eqref{eq:theorem} predicts that for $T>3T_{\rm RW}$, $C$ is spontaneously broken in the bulk of $A$, with $\Im L^{(3)}$ and $n_I$ as order parameters, while the bulk of $B$ remains $C$ symmetric.
Since $A$ and $B$ live on the same lattice and differ only in the numerator $p$, the difference between their bulks stems solely from the parity of $p+q$.
Consider also two nonrotating reference systems,
\begin{align}
A_0\colon\theta'=\pi,\qquad
B_0\colon\theta'=0,
\label{eq:reference}
\end{align}
with the same lattice action, lattice spacing and bare quark masses but with $3N_\tau$ temporal sites, so that their temperature is $1/(3N_\tau a)=T/3$.
If the lattice in the plane perpendicular to the axis and the action on it are invariant under the rotation, the boundary condition maps sites to sites, and the action terms linking the last and first time slices take the same form as on the nonrotating lattice.
The period argument above then carries over, and Eq.~\eqref{eq:theorem} predicts, without taking the continuum limit, that the bulk thermodynamic quantities of $A$ and $A_0$, and those of $B$ and $B_0$, coincide.
At finite lattice spacing, $T_{\rm RW}$ in the threshold is the RW endpoint temperature of the reference system with the same action and spacing.
This parameter-free prediction concerns not only the presence of a phase but the value of $n_I$ itself, independently of Eq.~\eqref{eq:nI}; moreover, the bare $\Im L^{(3)}$ of each rotating system and its reference system can be compared without determining the renormalization factor separately, since the closed path has the same length $3N_\tau a$ in both and the lattice action and spacing are the same.

In a finite volume with $C$-symmetric boundary conditions, $\langle\Im L^{(3)}\rangle=\langle n_I\rangle=0$, so spontaneous $C$ breaking is tested by two signatures: the probability distribution of $\Im L^{(3)}$ averaged over a region far from the axis develops one maximum at positive and one at negative values, and the susceptibility, the volume $V$ of the measurement region times the mean square of the region-averaged $\Im L^{(3)}$, grows in proportion to $V$~\cite{Bonati:2016pwz}.
The measurement region is chosen so that the correlation length of the equivalent system is well below both its distance from the axis and its distance from the edge of the lattice.
For $\tilde\Omega_I=2\pi/3$ and $4\pi/3$ the lattice in the plane perpendicular to the axis must be triangular; a triangular lattice is obtained from the ordinary cubic lattice by taking the rotation axis along its body diagonal, so that the standard Wilson fermion action can be used without modification (Appendix~\ref{app:lattice}).
The spinor rotation matrix in the quark boundary condition must be that of $4\pi/3$ for $B$; that of $-2\pi/3$, which moves the sites identically, would turn $B$ into a system equivalent to $A$, and Appendix~\ref{app:lattice} shows that the two differ only in the sign of the temporal boundary condition of the quark field.

\section{Discussion and outlook}
The $T/q$ correspondence and the parity rule in $p+q$ are known from thermal effective field theory~\cite{Benjamin:2024rvn}; what is obtained here is the identification of the parity rule with the imaginary quark chemical potential $\theta'$, Eq.~\eqref{eq:chemical}, the resulting set of $C$-breaking angular velocities with its threshold $qT_{\rm RW}$, Eq.~\eqref{eq:Cset}, together with the deconfinement and chiral thresholds, Eq.~\eqref{eq:Tc}, and the order parameter $L^{(q)}$ with the parameter-free lattice test.
Equations~\eqref{eq:theorem} and \eqref{eq:Cset} are exact once $T_{\rm RW}$ is supplied as nonperturbative input, and nonperturbative values of $\Im L^{(q)}$ and $n_I$ follow from lattice data of nonrotating QCD at the RW point evaluated at temperature $T/q$.
These bulk results cannot be analytically continued to real rotation, since the limit $r\to\infty$ at fixed $\tilde\Omega_I$ is nonanalytic in $\tilde\Omega_I$~\cite{Ambrus:2023bid}; they are statements about the Euclidean theory that lattice simulations address directly.

The remaining question is how the order parameters vary between the axis and the bulk.
On the axis the rotation leaves the point fixed, so the temperature remains $T$ and the rotation acts only on spin, and the minimum selected by the local effective potential there~\cite{Takada:2026rw} need not agree with the bulk state given by Eq.~\eqref{eq:theorem}.
For instance, $\tilde\Omega_I=2\pi/3$ lies outside the on-axis $C$-breaking interval $1.40\pi\lesssim\tilde\Omega_I\lesssim2.60\pi$, so the potential on the axis selects a $C$-symmetric minimum, whereas the bulk breaks $C$ for $T>3T_{\rm RW}$; conversely, at $\tilde\Omega_I=3\pi/2$ the potential on the axis selects a $C$-broken minimum, whereas the bulk is the equivalent system at $\theta'=0$ and is $C$ symmetric.
Likewise, the perturbative confinement found on the axis~\cite{Chen:2022smf,Chen:2024tkr} is a prediction of a one-loop calculation in the high-temperature limit, whereas at irrational $\tilde\Omega_I/2\pi$ the bulk is confined at every temperature, inheriting confinement from zero-temperature QCD.
The one-loop calculation on the axis is justified as $T\to\infty$, yet at $\tilde\Omega_I=2\pi/3$ and $3\pi/2$ the potential on the axis and the bulk select different states even in that limit.
The order parameter selected in the bulk may, however, reach the neighborhood of the axis through gradient terms and correlations; its actual value on the axis and its spatial distribution at finite radius $r$ require an analysis including gradient terms and are not addressed here.
At $T=qT_{\rm RW}$, with $r\to\infty$ taken first, the bulk inherits the criticality of the RW endpoint; how this criticality is approached at finite $r$, where the correlation length is not negligible compared with $r$, cannot be determined from Eq.~\eqref{eq:theorem} and remains, together with how the axis alters the critical fluctuations, a subject for future work.

\appendix

\section{Periods far from the rotation axis}
\label{app:periods}
Take a point $\bm x$ on the cylindrical surface at distance $r$ from the axis, and the local Cartesian coordinates $(\tau,s)$ with $s=r\varphi$ on that surface.
The boundary condition $(\tau,\varphi)\sim(\tau+\beta,\varphi-\tilde\Omega_I)$ and the $2\pi$ periodicity of $\varphi$ are represented by the two translations
\begin{align*}
v_1=(\beta,-\tilde\Omega_Ir),\qquad v_2=(0,2\pi r),
\end{align*}
and two points separated by $v_1$ or $v_2$ are the same point.
The set of all periods, the displacements that carry $\bm x$ to itself, is therefore
\begin{align*}
\Lambda=\{mv_1+nv_2\mid m,n\in\mathbb Z\}.
\end{align*}
For $\tilde\Omega_I=2\pi p/q$ with $\gcd(p,q)=1$, take integers $a,b$ with $qb-pa=1$ (B\'{e}zout's identity) and set
\begin{align}
w_1&=qv_1+pv_2=(q\beta,0),\notag\\[0.25em]
w_2&=av_1+bv_2=\biggl(a\beta,\dfrac{2\pi r}{q}\biggr)
\label{eq:lattice}
\end{align}
[Fig.~\ref{fig:geom}(a)].
Since $w_1$ and $w_2$ are integer linear combinations of $v_1$ and $v_2$, all their integer linear combinations belong to $\Lambda$; conversely, the matrix $\bigl(\begin{smallmatrix}q&p\\ a&b\end{smallmatrix}\bigr)$ mapping $(v_1,v_2)$ to $(w_1,w_2)$ has determinant $qb-pa=1$, so its inverse has integer entries, $v_1=bw_1-pw_2$ and $v_2=-aw_1+qw_2$, and every element of $\Lambda$ is an integer linear combination of $w_1$ and $w_2$.
Hence
\begin{align*}
\Lambda=\{kw_1+lw_2\mid k,l\in\mathbb Z\}.
\end{align*}
The translation $w_1$ is the pure imaginary-time translation of length $q\beta$ that leaves the spatial point $\bm x$ fixed, and its multiples exhaust the elements of $\Lambda$ whose $s$ component vanishes: the $s$ component of $mv_1+nv_2$ is $2\pi r(qn-pm)/q$, which vanishes only when $qn=pm$, that is, by $\gcd(p,q)=1$, when $(m,n)=k(q,p)$ and $mv_1+nv_2=kw_1$.
Every other period has an $s$ component of magnitude at least $2\pi r/q$.
The $s$ component is the arc length along the circle of radius $r$, and two points differing by a multiple of $2\pi r$ in $s$ are the same point; the periods $kw_1+lw_2$ with $l\equiv0\pmod q$ are of this kind, since $qw_2=aw_1+v_2$ gives $kw_1+lw_2=(k+al/q)w_1+(l/q)v_2$ with $l/q$ an integer: such a period is an integer combination of $w_1$ and $v_2$ and returns to the same spatial point as the multiples of $w_1$.
For $l\not\equiv0\pmod q$ the period $kw_1+lw_2$ has the $s$ component $2\pi rl/q$, so $\bm x$ and the point separated from it by this period lie on the same circle of radius $r$ at the central angle $2\pi l/q$; two points on a circle of radius $r$ at the central angle $\phi$ are at the distance $2r|\sin(\phi/2)|$, so their distance in three-dimensional space is the chord $2r|\sin(\pi l/q)|$.
Since $l$ and $l+q$ give the same point, we may take $l=1,\dots,q-1$; these $q-1$ points form, together with $\bm x$ itself, $q$ equally spaced points on the circle.
The angle $\pi l/q$ is one of $\pi/q,2\pi/q,\dots,(q-1)\pi/q$, and on this range $\sin$ is symmetric about $\pi/2$ and increasing from $0$ to $\pi/2$, so $|\sin(\pi l/q)|$ takes its minimum at the two ends $l=1$ and $l=q-1$, the two neighbors of $\bm x$, where it equals $\sin(\pi/q)$.
The distance is therefore at least $2r\sin(\pi/q)$ and grows without bound as $r\to\infty$.
By locality every such period drops out far from the axis; only the pure imaginary-time translations survive, and the bulk has the imaginary-time period $q\beta$.

\section{Finite radius near a rational angular velocity}
\label{app:window}
Equation~\eqref{eq:theorem} concerns the limit $r\to\infty$ at fixed $\tilde\Omega_I$.
Here we ask under what conditions the thermodynamics of the equivalent system at temperature $T/q$ is realized as the local thermodynamics at a finite radius $r$, that is, in a region around a point at radius $r$ small compared with $r$, when $\tilde\Omega_I=2\pi p/q+\delta$ with $\gcd(p,q)=1$, $q\geqslant2$ and $q|\delta|\ll1$.
Two kinds of periods bear on the local thermodynamics at radius $r$, and each gives one condition.

The first kind consists of the periods obtained by applying the boundary condition $k$ times with $1\leqslant k\leqslant q-1$.
Each relates the field at a point $\bm x$ at radius $r$ to the field at a point rotated by the angle $2\pi pk/q+k\delta$, whose distance from $\bm x$ is the chord $2r|\sin(\pi pk/q)|$ up to a correction of order $rk|\delta|\leqslant rq|\delta|$, hence at least $2r\sin(\pi/q)$ up to that correction.
By the locality stated in the text, the effect of a boundary condition on the local physics decays with the separation of the two points it relates on the scale of the correlation length $\xi$ of the equivalent system, so these periods cease to affect the local thermodynamics at radius $r$ once
\begin{align}
2r\sin(\pi/q)\gg\xi.
\label{eq:window}
\end{align}
This is the first condition; it is the same as for $\delta=0$ and gives a lower limit on $r$.

The second kind is the period obtained by applying the boundary condition $q$ times, $w_1$ of Appendix~\ref{app:periods}, which for $\delta\neq0$ becomes $(q\beta,-rq\delta)$ in the local coordinates: a translation by $q\beta$ in imaginary time combined with a spatial displacement of length $rq|\delta|$ along the circle.
On scales small compared with $r$ it imposes the boundary condition $\psi(\tau+q\beta,s-rq\delta)=(-1)^{p+q}\ee^{-\ii q\theta}\psi(\tau,s)$, that is, Eq.~\eqref{eq:qbc} with a spatial shift, and a shift does not become ineffective by locality, however small.
By the Euclidean rotation invariance of the local theory, a system with this boundary condition is in thermal equilibrium at the inverse temperature $\sqrt{(q\beta)^2+(rq\delta)^2}$~\cite{Giusti:2010bb}, so the local temperature at radius $r$ is
\begin{align}
T_{\rm loc}(r)\simeq\dfrac{T/q}{\sqrt{1+(rT\delta)^2}}.
\label{eq:Tloc}
\end{align}
The thermodynamics at temperature $T/q$ is realized only when the shift is negligible, $r|\delta|\ll\beta$; this is the second condition and gives an upper limit on $r$.
Here we have kept the leading order in $q\delta$, neglecting the residual spinor rotation by $q\delta$, which is $\bm1+\mathcal O(q\delta)$, so that the quark boundary phase is that of Eq.~\eqref{eq:qbc}; Eq.~\eqref{eq:Tloc} is a statement about the local physics at radius $r$, not an exact equivalence for the rotating system as a whole.
For $p=0$, $q=1$, where $\delta=\tilde\Omega_I$ and periods of the first kind do not exist, Eq.~\eqref{eq:Tloc} is the Tolman-Ehrenfest law at imaginary angular velocity, $T/\sqrt{1+\Omega_I^2r^2}$~\cite{Ambrus:2023bid}.

Together, the thermodynamics of the equivalent system at temperature $T/q$ is realized at radius $r$ in the window
\begin{align}
\dfrac{\xi}{2\sin(\pi/q)}\ll r\ll\dfrac{\beta}{|\delta|}.
\label{eq:windowB}
\end{align}
Beyond its upper limit the local temperature decreases smoothly with $r$ according to Eq.~\eqref{eq:Tloc}, as long as the same approximation remains valid.
Read at fixed $r$ as a condition on $q$, Eq.~\eqref{eq:window} fails for the rationals with $q\gg2\pi r/\xi$, with $\xi$ evaluated for each rational: for them the periods of the first kind relate points closer than the correlation length, and the locality argument above does not guarantee the thermodynamics at temperature $T/q$ at radius $r$.
At a finite radius, therefore, this argument establishes the structure only for the rationals with $q\ll2\pi r/\xi$, and the rationals of larger denominator join in as $r$ grows; the fractal structure of Fig.~\ref{fig:comb}(b) is complete only as $r\to\infty$.
For the lattice test of Eq.~\eqref{eq:pair}, $\delta=0$ by construction, so only Eq.~\eqref{eq:window} matters: the measurement region must lie farther from the axis than a few correlation lengths of the equivalent system, $r\gg\xi/\sqrt3$ for $q=3$, as stated in the text.
A systematic treatment of the finite-radius structure is left to a separate paper.

\section{Lattice implementation of the rotation}
\label{app:lattice}
For the boundary condition to map lattice sites to lattice sites, the spatial lattice must be invariant under the rotation by $2\pi p/q$ about the axis, so for $q=3$ its planes perpendicular to the axis must be triangular; a cubic lattice with the axis along a coordinate direction admits only $q=1,2,4$.
Take instead the cubic lattice with the axis along its body diagonal $(1,1,1)$.
The rotation by $2\pi/3$ about this axis is the cyclic permutation of the coordinates, $(x_1,x_2,x_3)\mapsto(x_3,x_1,x_2)$, which maps the cubic lattice onto itself, and the lattice planes perpendicular to the diagonal are triangular lattices: rotating the cubic lattice about its diagonal thus provides the rotating triangular lattice (this geometry is among the spatial tori compatible with a $2\pi/3$ rotation classified in Ref.~\cite{Nagy:2025rbx}).
Being a symmetry of the cubic lattice, the permutation maps sites to sites and links to links and leaves the standard gauge and Wilson fermion actions invariant; it is also a symmetry of the periodic box $(\mathbb Z_L)^3$, whose fixed points are the $L$ sites on the diagonal.
The boundary condition is therefore implemented as for the rotations by $\pi/2$ about a coordinate axis~\cite{Chernodub:2022veq}, by identifying the last time slice with the permuted first one, and the Wilson action can be used as it stands.

With the lattice fixed, we now write the quark boundary condition explicitly, which is where the spinor rotation enters.
Let $R(\varphi)$ denote the map of the sites under the rotation by $\varphi$ about the diagonal, so that $R(2\pi/3)$ is the permutation above, and let $S(\varphi)=\exp(-\ii\varphi\,\bm n\cdot\bm\varSigma/2)$ be the spinor matrix of the same rotation, with $\bm n=(1,1,1)/\sqrt3$, $\varSigma_k=\tfrac12\epsilon_{kij}\sigma^{ij}$, and $\sigma^{ij}=\tfrac{\ii}{2}[\gamma_i,\gamma_j]$ for Euclidean $\gamma$ matrices, $\{\gamma_i,\gamma_j\}=2\delta_{ij}$; its sense is fixed by $S(2\pi/3)^{-1}\gamma_1S(2\pi/3)=\gamma_2$ and cyclically.
The two properties used below are $R(\varphi+2\pi)=R(\varphi)$ and $S(\varphi+2\pi)=-S(\varphi)$, the spinor sign of a full turn.
The boundary condition $(\tau,\bm x)\sim(\tau+\beta,R(\tilde\Omega_I)^{-1}\bm x)$ of the text then reads for the quark field
\begin{align}
\psi(\tau+\beta,\bm x)=-\ee^{-\ii\theta}S(\tilde\Omega_I)\psi\bigl(\tau,R(\tilde\Omega_I)\bm x\bigr).
\label{eq:spinorbc}
\end{align}
At $\theta=0$, the systems $A$ ($\tilde\Omega_I=2\pi/3$) and $B$ ($4\pi/3$) of Eq.~\eqref{eq:pair} and, for comparison, the system $B'$ at $\tilde\Omega_I=-2\pi/3$ are defined by the three boundary conditions
\begin{align}
A\colon\ \psi(\tau+\beta,\bm x)&=-S(2\pi/3)\psi\bigl(\tau,R(2\pi/3)\bm x\bigr),\notag\\
B\colon\ \psi(\tau+\beta,\bm x)&=-S(4\pi/3)\psi\bigl(\tau,R(4\pi/3)\bm x\bigr)\notag\\
&=+S(-2\pi/3)\psi\bigl(\tau,R(-2\pi/3)\bm x\bigr),\notag\\
B'\colon\ \psi(\tau+\beta,\bm x)&=-S(-2\pi/3)\psi\bigl(\tau,R(-2\pi/3)\bm x\bigr),
\label{eq:ABbc}
\end{align}
where the second line of $B$ uses $R(4\pi/3)=R(-2\pi/3)$ and $S(4\pi/3)=-S(-2\pi/3)$.
The systems $B$ and $B'$ share the map of the sites $R(-2\pi/3)$ and the spinor matrix $S(-2\pi/3)$ and differ only in the sign on the right-hand side.
Implementing $B$ therefore requires nothing beyond the rotation by $-2\pi/3$ with the sign of the temporal boundary condition reversed, that is, a periodic instead of antiperiodic quark field; no spinor matrix of $4\pi/3$ has to be constructed.
That the sign matters is seen by applying the boundary condition three times, which closes the circuit of length $3\beta$: for $B$, $[S(-2\pi/3)]^3=-\bm1$, so the quark field is antiperiodic around the circuit and $\theta'=0$ by Eq.~\eqref{eq:chemical} with $p=2$, $q=3$; for $B'$, $[-S(-2\pi/3)]^3=+\bm1$, so it is periodic and $\theta'=\pi$ ($p=-1$, $q=3$), the same as for $A$, where $[-S(2\pi/3)]^3=+\bm1$.
Using the sign of $-2\pi/3$ for $B$ would thus simulate a system equivalent to $A$, as stated in the text.

\bibliography{main}

\end{document}